\documentstyle[preprint,aps]{revtex}

\begin{document}
\title{QUASI FREE $^{238}U(e,e^{\prime }f)$ -CROSS\ SECTION\ IN\
MACROSCOPIC-MICROSCOPIC\ APPROACH}
\author{V. P. Likhachev$^{a}$ J. Mesa$^{a,b}$ J. D. T. Arruda-Neto$^{a,c}$, B. V.\
Carlson$^{d}$, W.R. Carvalho$^{a}$ Jr, L.C. Chamon$^{a},$ M.T.F.da Cruz$^{a}$%
, A. Deppman$^{a}$, H. Dias$^{a}$, M. S. Hussein$^{a}$}
\address{$^{a}$ Instituto de F\'{i}sica, Universidade de S\~{a}o Paulo, S\~{a}o
Paulo, Brazil.\\
$^{b}$ Instituto Superior de Ciencias y Tecnologia Nucleares, Havana, Cuba.\\
$^{c}$ Universidade de Santo Amaro, Sao Paulo, Brazil.\\
$^{d}$ Instituto de Estudos Avan\c{c}ados-Centro T\'{e}cnico Aeroespacial, S%
\~{a}o Jos\'{e} dos Campos, Brazil.}
\maketitle

\begin{abstract}
We present the result of a theoretical study of inclusive quasi free
electrofission of $^{238}$U. The off-shell $\ $cross sections for the quasi
free reaction stage have been calculated within the Plane Wave Impulse
Approximation (PWIA), using a Macroscopic -Microscopic description of the
proton and neutron single particle momentum distributions. Electron wave
function distortion corrections were included \ using the effective momentum
approximation, and the Final State Interaction (FSI) effects were calculated
using an optical potential. The fissility for the proton single hole excited
states of the residual nucleus $^{237}$Pa was calculated \ both without and
with\ contributions of the pre-equilibrium emission of the particles. The
fissility for $^{237,238}U$ residual nuclei was calculated within the
compound nucleus model. The $(e,e^{\prime }f)-$cross sections thus obtained
were compared with available experimental data.
\end{abstract}


\section{INTRODUCTION}

A new aspect of investigations of quasi free scattering of high energy
electrons (QF)\ has been opened with the study of decay channels of single
hole states in the residual nucleus, created as a result of the QF process.
Especially interesting is to study fission following a QF process. In this
case we have an essentially single particle process in the first reaction
stage, and an essentially collective process in the final reaction stage.
The collective degrees of freedom are excited in the intermediate reaction
stage due to the residual interaction.

This is a new kind of nuclear reaction which may allow to get unique
information on the dissociation of well defined single hole configurations (
which we can select by coincidence\ $\ (e,e^{\prime }p)$) into complex
nuclear configurations, and its role in nuclear fission. \ In particular, to
study the limitations associated with the predictions of the shell model
based on the mean field approximation and residual forces for heavy nuclei
such as $^{238}$U.

The new and most important aspect of this reaction is that, after
knocking-out a proton or neutron, we obtain the heavy residual nucleus $%
^{237}$Pa or $^{237}$U in a single hole doorway state (see discussion
below), which could undergo nuclear fission. Indeed, instead of dealing with
collective doorway states, which are coherent sums of a great number of
1p-1h configurations ( a common situation, well-known giant resonances),
these non-collective doorway states will be represented by only one, well
defined, 1h configurations. The residual interaction mixes these 1h
configurations with more complicated 2h-1p and 3h-2p ones and fission may
occur either directly from 1h configurations, or, with some delay, from the
mixed states (or their components). Since in the case of QF we have in the
initial state only one configuration, the fission probability $P_{f}$ should
be more sensitive to the individual structure of this initial state as
compared with conventional reactions, where the effects of the structure are
averaged out over many single particle states forming the doorway. The
specific features of the quasi free reaction $^{238}$U $(e,e`pf)$ may create
a new situation when some of the single hole levels, which coincide in
energy position and quantum numbers with the collective levels in the second
well, have their fission probability enhanced and thus, the dependence on
state structure should be favored, which is also a rather unusual and
non-trivial situation.

The unambiguous extraction of single hole contributions is possible only in
an exclusive experimental scheme (reaction $(e,e^{\prime }$ $pf)$) This kind
of experiment involves extremely thin targets (fission fragments have to
leave the target with small energy losses), high energy resolution and
coincidence between the final particles (to separate the single hole states)
and has never been performed. The advent of high energy CW electron
accelerators, combined with the development of high resolution facilities,
opens the\ possibility of studying the fission channel for quasi free
electron scattering in an exclusive experimental setup. The most suitable
accelerator for this experiment is at the Thomas Jefferson National
Accelerator Facility (TJNAF).

Some integral properties of the quasi free electrofission could be studied
in inclusive experiments: $(e,f)$ \cite{Likha} and $(e,e^{\prime }f)$) \cite
{Hansen}. These works dealt only with the total issue of the QF contribution
in electrofission.

The goal of the present work is the calculation of the\ quasi free $%
(e,e^{\prime }f)$-differential cross section for $\ ^{238}$U, based on the
macroscopic-microscopic approach for the description of the quasi free
reaction stage, \ and the statistical theory for \ an estimate of the
fissility \ for single hole states in the residual nucleus. The comparison
with available \ inclusive experimental data will serve to check \ the
models \ used for the description of \ \ quasi free fission for heavy
deformed nuclei.

\section{PWIA SIX-FOLD DIFFERENTIAL CROSS SECTION}

In the first order Born approximation, the electron with initial
four-momentum $k_{1\mu }=(\overrightarrow{k}_{1},i\varepsilon _{1})$ and
final four-momentum\ $k_{2\mu }=(\overrightarrow{k}_{2},i\varepsilon _{2})$
interacts with the target nucleus and\ transfers\ \ a virtual photon with
four-momentum $q_{\mu }=(\overrightarrow{q},i\omega )=k_{1\mu }-k_{2\mu }$ \
\ ,\ \ leading to a final state with\ a knocked-out \ nucleon \ with $p_{\mu
}=(\overrightarrow{p},iE)$ \ and a residual nucleus with $P_{A-1\mu }=(%
\overrightarrow{P}_{A-1},iE_{A-1})$.

In the impulse approximation, \ a virtual photon interacts with a bound
nucleon (proton or neutron) of four-momentum $p_{m}=$\ $(\overrightarrow{p}%
_{m},iE_{m}),$which exits the nucleus with \ four-momentum $p_{\mu }=(%
\overrightarrow{p},iE)$ \ without further interaction (no FSI). In the PWIA
approximation\ $\overrightarrow{p}_{m}=-\overrightarrow{P}_{A-1}$ \ and the
missing quantities (momentum and energy of the nucleon before interaction)
can be defined from the energy and momentum conservation law in the
following way:

\begin{eqnarray}
\overrightarrow{p}_{m} &=&\overrightarrow{p}-\overrightarrow{q},  \label{1}
\\
E_{m} &=&\omega -T\ -T_{A-1},  \nonumber
\end{eqnarray}
where $\ E_{m}=M_{A-1}+m-M_{A}$\ is the nucleon missing (or separation)
energy , $T$ is the kinetic energy of the outgoing nucleon, and $T_{A-1}$ is
the kinetic energy of the residual nucleus. The momentum and energy of the
virtual photon can be varied independently.

In the plane wave impulse approximation (PWIA) the six-fold differential
cross section of the $(e,e^{\prime }p)-$ reaction in the Laboratory system
has the following form\cite{Forest83}: 
\begin{equation}
\frac{d^{6}\sigma }{d\Omega _{e}\text{ }d\Omega _{N}\text{ }d\varepsilon
_{2}dE\text{ }}=pE\sigma _{eN}S(E_{m},\overrightarrow{p_{m}}),  \label{2}
\end{equation}
\ \ \ \ \ \ \ \ \ \ \ \ \ where

\begin{equation}
\sigma _{eN}=\sigma _{mott}\left(
V_{C}W_{C}+V_{T}W_{T}+V_{I}W_{I}+V_{S}W_{S}\right)  \label{Eq. 3}
\end{equation}
\ is the off-shell \ electron-nucleon cross section, and $S(E_{m},%
\overrightarrow{p_{m}})$ \ is the spectral function which defines the
combined probability to find a bound nucleon with momentum $\overrightarrow{%
p_{m}}$ \ on the shell with separation energy $E_{m}.$

The kinematic functions $V$ in Eq. (\ref{Eq. 3})\ \ can be expressed ,
neglecting the mass of the electron, in the following way:

\begin{quotation}
\begin{eqnarray}
V_{C} &=&\frac{q_{_{\mu }}^{4}}{q^{4}}, \\
V_{T} &=&\frac{q_{_{\mu }}^{2}}{2q^{2}}+\tan ^{2}(\frac{\theta _{e}}{2}), \\
V_{I} &=&\frac{q_{_{\mu }}^{2}}{q^{2}}\cos \phi \sqrt{\frac{q_{_{\mu }}^{2}}{%
q^{2}}+\tan ^{2}(\frac{\theta _{e}}{2})}, \\
V_{S} &=&\frac{q_{_{\mu }}^{2}}{q^{2}}\cos ^{2}\phi +\tan ^{2}(\frac{\theta
_{e}}{2}),
\end{eqnarray}

and

\[
\sigma _{mott}=\frac{\alpha ^{2}\cos ^{2}\frac{\theta _{e}}{2}}{4\varepsilon
_{1}^{2}\sin ^{4}\frac{\theta _{e}}{2}}(1+\frac{2\varepsilon _{1}}{m_{p}}%
\sin ^{2}\frac{\theta _{e}}{2})^{-1}. 
\]
\end{quotation}

Above, $\sigma _{mott}$ is the Mott cross section, $\theta _{e}$ is the
electron scattering angle, and $\ \phi $ \ is the angle between the
scattering plane and the plane defined by the vectors \ $\overrightarrow{p}$
and $\ \overrightarrow{q},$ $\alpha =1/137$ is the fine structure constant.

For the structure functions W in Eq.(\ref{Eq. 3}) we use the off-shell
prescription of de Forest \cite{Forest83}:

\begin{eqnarray}
W_{C} &=&\frac{1}{4\bar{E}E}\{(\bar{E}+E)^{2}(F_{1}^{2}+\frac{\overline{q}%
_{\mu }^{2}}{4m^{2}}\kappa _{\text{ }}^{2}F_{2}^{2})-q^{2}(F_{1}+\kappa _{%
\text{ }}F_{2})^{2}\},  \label{211} \\
W_{T} &=&\frac{\overline{q}_{\mu }^{2}}{2\bar{E}E}(F_{1}+\kappa _{\text{ }%
}F_{2})^{2},  \nonumber \\
W_{I} &=&-\frac{p\sin \gamma }{\bar{E}E}(\bar{E}+E)(F_{1}^{2}+\frac{%
\overline{q}_{\mu }^{2}}{4m^{2}}\kappa _{\text{ }}^{2}F_{2}^{2}),  \nonumber
\\
W_{S} &=&\frac{p^{2}\sin ^{2}\gamma }{\bar{E}E}(F_{1}^{2}+\frac{\overline{q}%
_{\mu }^{2}}{4m^{2}}\kappa _{\text{ }}^{2}F_{2}^{2}),  \nonumber
\end{eqnarray}
where $\kappa _{\text{ }}$is the anomalous magnetic moment of the nucleon \
in units of the Bohr magneton \ ( $\kappa _{p}=$ 1.793 , $\kappa _{n\text{ }%
}=$ -1.913),

\begin{equation}
\bar{E}=\sqrt{p_{m}^{2}+m^{2}}\text{,}
\end{equation}
$m$ is the mass of the nucleon, $\overline{q}_{\mu }$ $=(\overrightarrow{q},i%
\overline{\omega })\ ,\ \overline{\omega }=E-\ \ \overline{E},$ $\gamma $\ \
is the angle between \ $\overrightarrow{p}$ and $\ \overrightarrow{q},$ and $%
\ F_{1}$ and $F_{2}$ are the on-shell Dirac and Pauli nucleon form factors,
respectively.

\begin{equation}
F_{1}(q_{_{\mu }}^{2})=\frac{1}{1+\frac{q_{_{\mu }}^{2}}{4m^{2}}}%
[G_{E}(q_{_{\mu }}^{2})+\frac{q_{_{\mu }}^{2}}{4m^{2}}G_{M}(q_{_{\mu
}}^{2})],
\end{equation}

\begin{equation}
\kappa _{p}F_{2}(q)=\frac{1}{1+\frac{q_{_{\mu }}^{2}}{4m^{2}}}%
[G_{M}(q_{_{\mu }}^{2})-G_{E}(q_{_{\mu }}^{2})],
\end{equation}
where\ for protons 
\begin{equation}
G_{E}^{p}(q_{_{\mu }}^{2})=(\frac{1}{1+\frac{q_{_{\mu }}^{2}}{0.71}})^{2},
\label{Eqn 13}
\end{equation}

\begin{equation}
G_{M}^{p}(q_{_{\mu }}^{2})=\mu _{p}G_{E}^{p}(q_{_{\mu }}^{2})],
\end{equation}
and \ for neutrons

\begin{equation}
G_{M}^{n}(q_{_{\mu }}^{2})=\mu _{n}G_{E}^{p}(q_{_{\mu }}^{2})],
\end{equation}

\begin{equation}
G_{E}^{n}(q_{_{\mu }}^{2})=\left| G_{M}^{n}(q_{_{\mu }}^{2})\right| \frac{%
q_{_{\mu }}^{2}}{4m^{2}}(\frac{1}{1+\frac{5.6q_{_{\mu }}^{2}}{4m^{2}}}),
\end{equation}
$\mu _{p}=1+\kappa _{p}=2.793$ and $\mu _{n}=\kappa _{n}=-1.913$ are the
proton and neutron magnetic moments \ in units of the Bohr magneton,
respectively, and $q_{_{\mu }}^{2}$ in Eq.(\ref{Eqn 13}) is in \ (GeV/c)$%
^{2};$

\section{PWIA THREE-FOLD DIFFERENTIAL CROSS SECTION}

In the independent particle shell model the spectral function for the
spherical orbitals $\alpha \equiv nlj$ with binding energy $E_{\alpha }$
takes the simple form:

\begin{equation}
S(E_{m},\overrightarrow{p_{m}})=\delta \left( E-E_{\alpha }\right) \text{ }%
\upsilon _{\alpha }^{2}\text{ }n_{\alpha }\left( \overrightarrow{p_{m}}%
\right) ,
\end{equation}
where $\upsilon _{\alpha }^{2}$ and $n_{\alpha }\left( \overrightarrow{p_{m}}%
\right) $are the occupation number and momentum distribution of the $\alpha $
orbital, respectively. The six-fold $(e,e^{\prime }p)-$cross section could
be transformed into a five-fold one:

\begin{equation}
\frac{d^{5}\sigma }{d\Omega _{e}\text{ }d\Omega _{N}\text{ }dE}=pE\sigma
_{eN}\ \upsilon _{\alpha }^{2}n_{\alpha }\left( \overrightarrow{p_{m}}%
\right) ,  \label{f}
\end{equation}
where energy and momentum conservation are imposed for the kinematic
variables that appear\ in $\sigma _{eN}$

To obtain the three-fold $\ (e,e^{\prime })-$cross sections \ for each bound
proton and neutron orbital we have \ integrated Eq.(\ref{f}) over $d\Omega
_{N}$ $\ \ $using a Monte Carlo approach. For each fixed $\ \varepsilon _{2}$
we calculate $q$ \ neglecting the recoil energy of the residual nucleus:

\begin{equation}
q=\sqrt{k_{1}^{2}+k_{2}^{2}-2k_{1}k_{2}\cos \theta _{e}},
\end{equation}
\ \ \ \ \ \ and the kinetic \ energy of ejected nucleon:

\begin{equation}
T=\omega -E_{\alpha }.
\end{equation}

Then, we generate \ randomly and uniformly the directions \ of the ejected
nucleon with respect to $\overrightarrow{q}$ \ and calculate the
corresponding momentum of the internuclear nucleon $p_{m}$. Finally, using
the values of $\upsilon _{\alpha }^{2}$ \ and\ $n_{\alpha }\left(
p_{m}\right) $ (see below) we calculate

\begin{equation}
\frac{d^{5}\sigma }{d\Omega _{e}\text{ }d\Omega _{N}\text{ }d\varepsilon _{2}%
}(p_{m})
\end{equation}

The three-fold $\ (e,e^{\prime })-$cross section \ was obtained as:

\begin{equation}
\frac{d^{3}\sigma }{d\Omega _{e}\text{ }d\varepsilon _{2}\text{ }}%
(\varepsilon _{2})=\text{ }<\frac{d^{5}\sigma }{d\Omega _{e}\text{ }d\Omega
_{N}\text{ }d\varepsilon _{2}}>4\pi
\end{equation}
where the average five-fold cross section is

\begin{equation}
<\frac{d^{5}\sigma }{d\Omega _{e}\text{ }d\Omega _{N}\text{ }d\varepsilon
_{2}\text{ }}>=\frac{\sum \frac{d^{5}\sigma }{d\Omega _{e}\text{ }d\Omega
_{N}\text{ }dE_{N}\text{ }}}{N},
\end{equation}
and $N$ \ is the number of Monte Carlo starts

\section{SINGLE PARTICLE BOUND STATES}

The single particle bound state energies and momentum distributions were
calculated in the framework of the macroscopic-microscopic approach using
the BARRIER code \cite{Barrier99}.

The energy of the nucleus is given as:

\begin{equation}
E_{tot}=E_{LD}+\delta E_{shell},  \label{111}
\end{equation}
where $E_{LD}$ $\ \ $is the macroscopic liquid drop \ part of the energy and 
$\delta E_{shell}$\ \ is\ the shell correction, which describes the shell
and pairing effects. Both shell correction and the macroscopic part of the
energy have been calculated according to \cite{Barrier99}.

Axially symmetric nuclear shapes have been considered in the present work,
and from these potential energy surfaces, the equilibrium (ground state)
deformation parameters ${\varepsilon }$ (elongation) and ${\alpha _{4}}$
(hexadecapolar momentum) have been calculated by minimizing the total
nuclear energy ( Eq.(\ref{111})), obtaining thus, $\varepsilon =0.227$ and $%
\alpha _{4}=0.059$.

An Woods- Saxon potential\cite{WoSa99} consisting of a central part $V$,
spin-orbit part $V_{SO}$, and Coulomb part $V_{Coul}$ for protons was
employed:

\begin{equation}
V^{WS}(r,z,\varepsilon ,\widehat{\alpha })=V(r,z,\varepsilon ,\widehat{%
\alpha })+V_{so}(r,z,\varepsilon ,\widehat{\alpha })+V_{Coul}(r,z,%
\varepsilon ,\widehat{\alpha })
\end{equation}
The real potential $V(r,z,\varepsilon ,\widehat{\alpha })$ involves the
parameters V$_{0}$, $r_{0}$ and $a$ , describing the depth, radius and
diffuseness of the central potential, respectively, and is expressed as:

\begin{equation}
V(r,z,\varepsilon ,\widehat{\alpha })=\frac{V_{0}}{1+exp\left[ \frac{%
dist(r,z,\varepsilon ,\widehat{\alpha })}{a}\right] },
\end{equation}
where $dist(r,z,\varepsilon ,\widehat{\alpha })$ is the distance between a
point and the nuclear surface, and $\varepsilon $ and $\widehat{\alpha }$
are deformation parameters.

The depth of the central potential is parametrized as

\begin{equation}
V_{0}=V_{0}[1\pm \kappa (N-Z)/(N+Z)],
\end{equation}
with the plus sign for protons and the minus sign for neutrons, with the
constant $\kappa =0.63$.

The spin-orbit interaction is then given by:

\begin{equation}
V_{so}(r,z,\varepsilon ,\widehat{\alpha })=\lambda \left( \frac{h}{2Mc}%
\right) ^{2}\nabla V(r,z,\varepsilon ,\widehat{\alpha })\cdot (\vec{\sigma}%
\times \vec{p}),
\end{equation}
where $\lambda $ denotes the strength of the spin--orbit potential and M is
the nucleon mass. The vector operator $\vec{\sigma}$ stands for Pauli
matrices and $\vec{p}$ is the linear momentum operator.

The Coulomb potential is assumed to be the one corresponding to the nuclear
charge $(Z-1)e$, uniformly distributed inside the nucleus.

For the ground state deformation of $^{238}$U, small changes in $\lambda $
(spin-orbit potential strength) and $r_{0-so}$ (spin-orbit potential radius)
of the Chepurnov parameters\cite{Chepurnov68} are introduced in order to
reproduce adequately the spin/parity of the levels sequence. Using single
particle states obtained by this procedure, the quasiparticle states can be
calculated for the first minimum region, providing spin, parity, energy and
level spacing for the ground and some low--lying states. The quasiparticle
spectrum was obtained by using the semi-microscopic combined method \cite
{Dencom95}.

The potential parameters were chosen to give the best fit to the spectrum of
single-quasiparticle excitations of the neighboring nuclei. The Hamiltonian
matrix elements are calculated with the wave functions of a deformed axially
symmetric oscillator\ potential. The wave functions $\phi _{i}$ in the
coordinate space are expanded into eigenfunctions of the axially deformed
harmonic oscillator potential. From this expansion, we conveniently express
the single particle Woods-Saxon wave function in momentum space. These
single particle momentum distributions were averaged over the nuclear
symmetry axis directions.The occupation probabilities were calculated in the
framework of the BCS model \cite{Dencom95}. The results for the occupation
number calculations are shown in Fig. 1. Fig. 2 shows 6 typical averaged
momentum distribution for proton bound states.

\section{FISSILITY}

The quasi free knockout of nucleons leads to the excitation of the residual
nucleus \ $^{237}Pa,$ in the case of \ $\ (e,e^{\prime }p)$ reaction, and $%
^{237}U$ \ for $\ (e,e^{\prime }n).$The excitation energy $E^{\ast }$%
(nucleus A-1) has two origins: holes in the shells of the nucleus A, which
appear as a result of the knockout of nucleons, and final state interaction
(FSI) of the outgoing nucleon.

The fast, quasi free \ reaction stage occurs \ at zero thermal excitation
(ground state) of the initial nucleus $^{238}U,$ and results in a single
hole in one of the shells. This single hole configuration \ \ forms a
doorway for a thermalization process which leads to the thermal excitation $%
E^{\ast }$of the residual nucleus $.$

The thermalization is a complicate process which involves creation of new \
many particle-hole configurations in competition with particle emission and
fission, and for some doorway configurations it might have non-statistical
character, but, as a first guide-line for order of magnitude estimates we \
calculate the total fission probability (nucleus with energy $E^{\ast }$
deexcites in several steps) using the statistical \ theory, both \ with \
and without taking into account the preequilibrium \ decay.

\subsubsection{ Thermalization without preequilibrium decay.}

Firstly, we considered an extreme situation, by assuming that the residual
interaction leads to thermalization and formation \ of compound nucleus just
after the fast reaction stage, without any preequilibrium particle emission.
In this case, the compound nucleus excitation energy is assumed to be :

\begin{equation}
\ E^{\ast }=-E_{\alpha }+E_{f}  \label{Eq. 29}
\end{equation}

where $E_{\alpha }$ and $E_{f}$\ \ are the energies of the bound state
(hole) and Fermi level, respectively.

For the calculations of the compound nucleus fissility \ we used \ the
Bohr-Wheeler \cite{Bohr} and Weisskopf \cite{Weisskopf}\ models\ for the
description of the evaporation/fission competition. A Monte Carlo algorithm 
\cite{Deppman} was developed for the evaporation/fission process which
includes not only the neutron evaporation vs fission competition, but also
takes into account the proton and alpha-particle contributions.

The Monte Carlo code for Evaporation-Fission \cite{Deppman1} calculates, at
the \ i$^{th}$ step of the evaporation chain, the fission probability $F_{i}$%
, and then chooses randomly which particle will evaporate (neutron, proton
or alpha particle), according to their relative branching ratios. After
evaporation, the mass, atomic number and excitation energy \ of the new
residual nucleus are calculated. This process continues until the excitation
energy available in the nucleus is not enough to emit any of the possible
evaporating particles. At this point, the evaporation process stops, and we
can calculate the nuclear fissility by the expression

\begin{equation}
W=\sum_{i}\left[ \prod_{j=0}^{i-1}(1-F_{j})\right] F_{i},
\end{equation}

where $\ $

\begin{equation}
F_{j}=\frac{(\frac{\Gamma _{f}}{\Gamma _{n}})_{i}}{1+(\frac{\Gamma _{f}}{%
\Gamma _{n}})_{i}+(\frac{\Gamma _{p}}{\Gamma _{n}})_{i}+(\frac{\Gamma
_{\alpha }}{\Gamma _{n}})_{i}},
\end{equation}

$\ \Gamma _{f},\Gamma _{n},\Gamma _{p},\Gamma _{\alpha }$ are fission,
neutron, proton and alpha-particle decay widths, respectively.

The probability for the emission of a particle $j$ is calculated within the
Weisskopf statistical model\cite{Weisskopf} . The level density for the
initial and final nucleus are calculated from the Fermi gas expression.

The fission barriers were taken as $6.13$, $5.2$ and $5.62$ MeV\ and the
neutron separation energies as $\ 5.78,5.12$ and $5.04$ MeV in the first
step for $^{237}Pa$ $^{237}U,$ and \ $^{238}U$ \ , respectively. These
values reproduce the experimental data for P$_{f}$ (see below).

For the other steps in the evaporation chain the fission barrier is
calculated as\cite{Tavares},

\begin{equation}
\ B_{f}=C(0.22(A-Z)-1.40Z+101.5)\text{ MeV,}  \label{25}
\end{equation}

where $\ \ C=1-\frac{E^{\ast }}{B}$ is the factor which take into account
the dynamical effects \cite{Tavares}, \ $B$ is the total nuclear binding
energy\cite{Tavares}, and $E^{\ast }$ is the nuclear excitation energy.

The neutron separation energy was taken from \cite{Guaraldo} for the other
steps.

Using the model described above, we have calculated the fissility for\ $%
^{237}Pa,$ $^{237}U$ and \ $^{238}U$\ \ \ (figure 3, solid curves). The
peaks observed in the fissility correspond to the opening of the fission
channel in the daughter nuclei. In figure 3 the experimental data for the
fissility of $^{237}Pa$ , $^{237}U$ ( \cite{Vandenbosch}) and $^{238}U$ ( 
\cite{Vandenbosch},\cite{Hansen}) are also shown \ by dark strips. Data for $%
^{237}Pa$ were obtained by extrapolation of the neutron to fission width
ratios for $Z=91$ and $A=230,231,232,233$ \cite{Gavron} to $A=237$ , using
the empirical trend presented in Vandenbosch and Huizenga \cite{Vandenbosch}.

It should be pointed out that in our calculations of \ the fissility we
assumed that the hole excitation energies for an A-1 nucleus correspond to
the compound nucleus excitation energies (see Eq. (\ref{Eq. 29})), that is
to say, complete thermalization is reached without any preequilibrium decay.
Such calculations could be considered as upper limits for the fissility.

\subsubsection{\ Account of the preequilibrium\ decay}

The calculation of preequilibrium decay \cite{Bonetti}, \cite{Gadioli} was
performed \ only for $^{237}Pa,$ since the $(e,e^{\prime }p)$-reaction gives
the main contribution to the \ total QF cross section.\ We used the exciton
model{\ \cite{Griffin}, \cite{Cline}, \cite{Braga}} and the code STAPRE. In
this model, the states of the system are classified according to the number
of excitons $n$, which corresponds to the total number of excited particle $%
p $ and hole $h$ degrees of freedom, $n=p+h$. Starting from a simple
configuration of low exciton number, the system is assumed to equilibrate
through a series of two-body collisions and to emit particles from all
intermediate states. The application of a two-body interaction to states of
a $(p,h)$ configuration results in states with $(p+1,h+1)$, $(p,h)$, and $%
(p-1,h-1)$ excited particles and holes. The transition rates, which are
taken to be averages over all states of a configuration, do depend on the
number of excited particles and holes, and $\lambda _{+}(n)$ are the average
rates for internal transitions from the $n$ exciton configuration with a
change of exciton number by -2, 0, or +2.

The decay is described using the Hauser-Feshbach formalism. We have
considered fission in competition with neutron and gamma emission.

The initial configuration in $^{237}$Pa, consistent with the proton knockout
reaction for $^{238}U$ \ initiating the statistical cascade,\ consists of
one-particle at the Fermi level and one-hole in a bound state. Our
calculations were performed \ assuming only an one-hole initial
configuration \ of \ the l=0 partial wave alone. The particle at the \ Fermi
level contributes negligibly to the equilibration process. The fission
barriers, neutron separation energies \ and level density parameters were
taken to be the same as those of the \ compound nucleus calculations in the
previous section.

The exciton model fissility results for single hole states of $^{237}$Pa \
are shown in fig 3 by the dotted curve. We note \ that these calculations \
for fissility show a \ smoother behavior than that for the compound model.
The preequilibrium \ particle emission \ removes \ some excitation energy \
before a equilibrium is reached reducing, therefore, the probability of
opening new \ chances for fission.

\bigskip

\section{DISTORTION CORRECTIONS TO PWIA}

\bigskip

An exact treatment \ of distortions of \ \ initial and final electrons and
knocked-out nucleon waves,\ arising from the vicinity of a heavy nucleus,
requires the solution of \ the Dirac equation for a large number of partial
waves \ and is a hard \ special task. In the case of an inclusive quasi free
reaction, \ the distortion corrections \ are averaged over directions of
ejected nucleon and we will treat these distortions \ qualitatively in an
effective way.

\bigskip

The main effect of electron wave distortions, produced by the attractive
Coulomb potential of \ a nucleus, could be treated as an effective increase
of \ the momentum transfer (effective momentum approximation (EMA) \cite
{Boffi}). For the calculation of $q_{eff}$ we use the prescription of \cite
{Giusti}, namely

\[
\overrightarrow{q_{eff}}=\overrightarrow{q}(1+\frac{3Z\alpha }{2\varepsilon
_{1}R})-\frac{3Z\alpha }{2\varepsilon _{1}R}\omega \frac{\overrightarrow{%
k_{2}}}{k_{2}}, 
\]
where $R=1.2A^{1/3}$ is the equivalent nuclear radius. \ The account of the\
Coulomb distortions in EMA reduces the cross section since it goes roughly
as $q_{eff}^{-4}.$

\bigskip

The\ distortions arising from FSI of \ an ejected nucleon \ with the
residual nucleus was included in an effective way \cite{Kenzo} assuming that
\ the nucleon, propagating as a free particle, sees the energy dependent
optical potential V, where \ \ V \ is \ the average depth \ of the real part
of the optical potential in the region of space in which the initial bound
state wave functions are not small (r%
\mbox{$<$}%
8 fm), and \ as \ a result \ carries an effective local momentum $p_{eff}$

\bigskip

We have used a phenomenological optical potential with significant
dependence on the bombarding energy which is a result of the scattering data
analysis over a large energy range\cite{Brandan}. The theoretical model to
account for this energy-dependence is based on nonlocal quantum exchange
effects \cite{Chamon1},\cite{Ribeiro}. Within this model, the nuclear
interaction $V_{N}$ is connected with the folding potential $V_{F}$ through 
\begin{equation}
V_{N}(R,E)\approx V_{F}(R)\;e^{-4v^{2}/c^{2}}\;,
\end{equation}
where $c$ is the speed of light and $v$ is the local relative speed between
the two partners (proton/neutron and residual nucleus), and 
\begin{equation}
v^{2}(R,E)=\frac{2}{\mu }\left[ E-V_{C}(R)-V_{N}(R,E)\right] \;.
\end{equation}

The folding potential depends on the densities of the two partners in the
collision 
\begin{equation}
V_{F}(R)=\int \rho _{1}(r_{1})\;\rho _{2}(r_{2})\;u_{0}(\vec{R}-\vec{r_{1}}+%
\vec{r_{2}})\;d\vec{r_{1}}\;d\vec{r_{2}}\;.
\end{equation}
The standard M3Y interaction ``frozen'' at 10 MeV/nucleon \cite{Chamon2} has
been assumed for the effective nucleon-nucleon interaction, $u_{0}(\vec{r})$.

Fig.4 shows the average potential depth ($<V_{n}>)$ as a function of \ the
energy for protons (residual nucleus $^{237}Pa)$ and neutrons ($^{237}U).$
This average depth effectively incorporates distortions connected with FSI \
and modifies the asymptotic nucleon kinetic energy:

The effect of the optical potential was introduced \ by replacing the
momentum of the ejected nucleon by an effective momentum\cite{Kenzo},\cite
{Klavansky}:

\begin{equation}
p_{eff}=p\sqrt{1-\frac{<V_{n}>}{T}}\text{ \ },
\end{equation}
where $\ T$ is the kinetic energy of the ejected nucleon without FSI. This
procedure was included in the Monte Carlo simulations.

We did not take into account \ distortions \ arising from \ a focusing of
electron waves \ and from the\ imaginary part of the optical potential,
since these effects are likely to\ compensate each other.

\section{ DISTORTION CORRECTIONS TO FISSILITY}

The fissility $P_{f}$ for each proton and neutron \ bound state was
calculated \ without and with \ the effect of FSI.

In the calculation without FSI we assume that the excitation energy of the
doorway states is: $E^{\ast }=-E_{\alpha }+E_{F},$ \ where $E_{\alpha }$ and 
$E_{F}$\ are the nucleon binding energies for shell $\alpha $ and Fermi
level, respectively (see tables 1,2)\ .To estimate \ the energy deposited \
by the ejected nucleon in the residual nucleus as a result of \ the FSI, we
assumed that the losses of the nucleon kinetic energy ($\Delta T),$ \
resulting from its passage\ through \ the imaginary part of the optical
potential, \ are deposited in the residual nucleus \ and the excitation
energy of the doorway states is $E_{FSI}^{\ast }=-E_{\alpha }+E_{F}+\Delta
T, $ where $\Delta T=(1-\exp (\frac{<W>\Delta t}{\hbar }))\frac{T}{2},$ $<W>$
\ is the average depth of the imaginary part of the optical potential, which
was taken from the systematics \cite{Becchetti}, and $\Delta t$ is the
average nucleon flight time through the imaginary part of the optical
potential.

This procedure to take into account the FSI was included in the Monte Carlo
simulations. For each fixed \ transferred $\ \ \omega $ \ we \ choose
randomly proton or neutron bound state numbers, and calculate $%
\overrightarrow{q}$ \ neglecting the recoil energy of the residual nucleus.
Then, we generate \ uniformly the directions \ of the ejected nucleon with
respect to $\overrightarrow{q},$ define the kinetic energy of the outgoing
nucleon $T$ , and finally calculate $E_{FSI}^{\ast }$ , the corresponding
fissility, and the\ total $(e,e^{\prime }f)$-cross section (see Eq.(\ref{60}%
))

\section{FINAL RESULTS}

To check how physically reasonable the microscopic-macroscopic approach used
for the description of QF\ fission is, calculations were done with the
kinematics conditions of \cite{Hansen}, which is the only available
experimental \ data \ for the reaction \ under study.

Fig.5 shows the three fold \ quasi free \ PWIA\ cross sections for four
proton orbitals of $^{238}$U at $\varepsilon _{1}=750$ MeV and $\theta
e=37.5^{0}$. It is seen that the cross sections show different spectral
shapes, which is the result of differences in the momentum distributions for
these orbitals. The maxima are shifted with respect to each other according
to the differences in the separation energies.

The total $(e,e^{\prime })-$cross section is the incoherent sum of the
contributions from all proton and neutron orbitals.

Fig.6 shows the total quasi free $(e,e^{\prime })-$ cross section for $%
\varepsilon _{1}=750$ MeV and $\theta e=37.5^{0}$. The upper part shows the
results of the calculations in PWIA, that is, without any corrections
arising from the distortions of electron \ and nucleon waves \ in the
vicinity of the nucleus, while in the lower \ part the distortions were
considered (substitution \ $q\ \Rightarrow \ q_{eff\text{ }},$ $p$ $%
\Rightarrow $ $p_{eff}$ ). The dark circles show the proton contributions,
dark triangles - neutron, and light circles- total cross section. The
hatched area shows the experimental data \cite{Hansen} after extraction of
the $\Delta $ resonance peak. It is seen that optical/mean-field distortions
\ significantly reduce the cross section and slightly shift it to a lower $%
\omega .$ The calculations with distortions qualitatively \ reproduce the
experimental data without \ any fitting parameters. All parameters used in
the calculations were fixed \ at the microscopic stage.

The differential cross section of the\ $(e,e^{\prime }f)$-reaction was
obtained \ for each proton $(p)$ and neutron $(n)$ bound state $\alpha ,$ \
assuming an isotropic angular distribution for the fission fragments and \
taking the fissility of the \ residual nucleus as a factor $P_{f}$ :

\begin{equation}
\frac{d^{5}\sigma _{p,n,\alpha }}{d\Omega _{e}\text{ }d\varepsilon _{2}d%
\text{ }\Omega _{f}}(\varepsilon _{2})=\frac{1}{4\pi }\frac{d^{3}\sigma
_{_{p,n,\alpha }}}{d\Omega _{e}\text{ }d\varepsilon _{2}}(\varepsilon
_{2})P_{f}  \label{60}
\end{equation}

The total inclusive quasi free $(e,e^{\prime }f)-$ cross section was
obtained as a sum of proton and neutron contributions:

\[
\frac{d^{5}\sigma _{tot}}{d\Omega _{e}\text{ }d\varepsilon _{2}d\text{ }%
\Omega _{f}}(\varepsilon _{2})=\sum \frac{d^{5}\sigma _{p,a}}{d\Omega _{e}%
\text{ }d\varepsilon _{2}d\text{ }\Omega _{f}}(\varepsilon _{2})+\sum \frac{%
d^{5}\sigma _{n,a}}{d\Omega _{e}\text{ }d\varepsilon _{2}d\text{ }\Omega _{f}%
}(\varepsilon _{2})\text{ .} 
\]

Fig. 7 shows the cross section for $\varepsilon _{1}=750$ MeV and $\theta
e=37.5^{0}.$ The upper part shows \ the calculation in PWIA with distortion
corrections to the cross section (by using $q_{eff},$ $p_{eff}$) ,\ but
without correction of \ the fissility arising from additional excitation
(FSI) of the residual nucleus. The lower part shows results similar to the
upper part but with FSI corrections \ to fissility. Light circles in both
parts of the fig. 7 show the results of calculations which take into account
the preequilibrium emission, dark circles - without accounting of\ the
preequilibrium emission, and the dark triangle with error bars show the
experimental data\cite{Hansen}. One sees a strong influence of FSI fissility
corrections on the $(e,e^{\prime }f)-$ cross section, and the calculations \
with this effect taken into account agree with experimental data. The
calculations without and with preequilibrium emission \ gave close results.
Inclusive cross sections are not sensitive to the difference in the reaction
mechanisms.

\section{CONCLUSIONS}

We presented in this paper a theoretical study of the inclusive quasi free
electrofission of $^{238}U.$ The proton and neutron bound state
characteristics were calculated in the framework of the
macroscopic-microscopic approach, using the axially deformed Woods-Saxon
single particle potential. The occupation numbers were calculated in the BCS
approach. The differential cross sections for the quasi free scattering
reaction stage were calculated in PWIA, using off-shell electron-nucleon
cross section corrected for Coulomb distortions of electron waves, and FSI
distortions for ejected nucleons. The corrected\ quasi free cross sections
reproduce the experimental data fairly well.

The fissility for the single hole states of the residual nuclei was
calculated \ with and \ without FSI \ in the framework of two approaches:
compound nucleus model without the preequilibrium emission of particles, and
the exciton model, which allows for the preequilibrium emission of
particles. The account of \ the FSI corrections to fissility \ strongly
increases it and \ the calculations agree with available experimental data
for the quasi free electrofission of $^{238}U.$

The calculations with and without the preequilibrium emission\ give close
results, showing that in inclusive cross sections the effect of
preequilibrium emission, averaged over all shells, become small.

In conclusion, the approach based on the microscopic-macroscopic description
of nucleon bound states, in conjunction with the plane wave impulse
approximation and mean field distortions, \ give adequate description of
inclusive quasi free electrofission at GeV energy range.

\newpage

\section{ACKNOWLEDGMENT}

\bigskip

The authors thank the Brazilian agencies CNPq and FAPESP for the partial
support to this work, and the graduate student M.S. Vaudeluci for the help
in the development of the Monte Carlo code.

\newpage

\section{Table 1}

\[
\text{%
\begin{tabular}{|l|l|l|l|l|l|l|l|l|l|l|l|}
\hline
n{\small \ } & {\small \ \ [MeV]} & {\small \ \ }$\pi ${\small J} & {\small %
\ \ \ \ \ [N n}$_{z}${\small \ }$\Lambda ${\small ]} & n & {\small \ [MeV]}
& {\small \ \ }$\pi ${\small J} & {\small \ \ \ \ \ [N n}$_{z}${\small \ }$%
\Lambda ${\small ]} & n & {\small [MeV]} & {\small \ \ }$\pi ${\small J} & 
{\small \ \ \ \ \ [N n}$_{z}${\small \ }$\Lambda ${\small ]} \\ \hline
{\small 1} & {\small -33.685} & {\small \ 1/2 } & {\small 1/2 [ 0 0 0]} & 
{\small 23} & {\small -16.192 } & {\small -3/2 } & {\small 3/2 [ 3 0 1]} & 
{\small 45} & {\small -7.491 } & {\small \ 3/2 } & {\small 3/2 [ 4 0 2]} \\ 
\hline
{\small 2} & {\small -31.397 } & {\small -1/2 } & {\small 1/2 [ 1 1 0]} & 
{\small 24} & {\small -15.490 } & {\small -1/2 } & {\small 1/2 [ 3 0 1]} & 
{\small 46} & {\small -7.195 } & {\small \ 1/2 } & {\small 1/2 [ 4 0 0]} \\ 
\hline
{\small 3} & {\small -30.043 } & {\small -3/2 } & {\small 3/2 [ 1 0 1]} & 
{\small 25} & {\small -15.415 } & {\small \ \ 7/2 } & {\small 7/2 [ 4 1 3]}
& {\small 47} & {\small -6.277 } & {\small \ 5/2 } & {\small 5/2 [ 6 4 2]}
\\ \hline
{\small 4} & {\small -29.670 } & {\small -1/2 } & {\small 1/2 [ 1 0 1]} & 
{\small 26} & {\small -14.529 } & {\small \ \ 9/2 } & {\small 9/2 [ 4 0 4]}
& {\small 48} & {\small -6.189 } & {\small -5/2 } & {\small 5/2 [ 5 2 3]} \\ 
\hline
{\small 5} & {\small -28.141 } & {\small \ \ 1/2 } & {\small 1/2 [ 2 2 0]} & 
{\small 27} & {\small -14.302 } & {\small \ \ 3/2 } & {\small 3/2 [ 4 2 2]}
& {\small 49} & {\small -5.348 } & {\small -3/2 } & {\small 3/2 [ 5 2 1]} \\ 
\hline
{\small 6} & {\small -26.630 } & {\small \ \ 3/2 } & {\small 3/2 [ 2 1 1]} & 
{\small 28} & {\small -13.984 } & {\small -1/2 } & {\small 1/2 [ 5 3 0]} & 
{\small 50} & {\small -4.827 } & {\small \ 7/2 } & {\small 7/2 [ 6 3 3]} \\ 
\hline
{\small 7} & {\small -25.963 } & {\small \ \ 1/2 } & {\small 1/2 [ 2 1 1]} & 
{\small 29} & {\small -13.111 } & {\small \ \ 1/2 } & {\small 1/2 [ 4 2 0]}
& {\small 51} & {\small -4.340 } & {\small -7/2 } & {\small 7/2 [ 5 1 4]} \\ 
\hline
{\small 8} & {\small -25.542 } & {\small \ \ 5/2 } & {\small 5/2 [ 2 0 2]} & 
{\small 30} & {\small -13.091 } & {\small -3/2 } & {\small 3/2 [ 5 4 1]} & 
{\small 52} & {\small \ -3.949 } & {\small -1/2 } & {\small 1/2 [ 5 2 1]} \\ 
\hline
{\small 9} & {\small -24.473 } & {\small \ \ 3/2 } & {\small 3/2 [ 2 0 2]} & 
{\small 31} & {\small -12.383 } & {\small \ \ 5/2 } & {\small 5/2 [ 4 1 3]}
& {\small 53} & {\small -3.667 } & {\small \ \ 1/2 } & {\small 1/2 [ 6 5 1]}
\\ \hline
{\small 10} & {\small -24.025 } & {\small \ -1/2 } & {\small 1/2 [ 3 3 0]} & 
{\small 32} & {\small -11.735 } & {\small \ -5/2 } & {\small 5/2 [ 5 3 2]} & 
{\small 54} & {\small -3.465 } & {\small -5/2 } & {\small 5/2 [ 5 1 2]} \\ 
\hline
{\small 11} & {\small -22.836 } & {\small \ \ 1/2 } & {\small 1/2 [ 2 0 0]}
& {\small 33} & {\small -11.053 } & {\small \ \ 7/2 } & {\small 7/2 [ 4 0 4]}
& {\small 55} & {\small -3.417 } & {\small \ 9/2 } & {\small 9/2 [ 6 2 4]}
\\ \hline
{\small 12} & {\small -22.716} & {\small \ -3/2} & {\small 3/2 [ 3 2 1]} & 
{\small 34} & {\small -10.831 } & {\small \ \ 3/2 } & {\small 3/2 [ 4 1 1]}
& {\small 56} & {\small -3.051 } & {\small -9/2 } & {\small 9/2 [ 5 0 5]} \\ 
\hline
{\small 13} & {\small -21.614 } & {\small -1/2 } & {\small 1/2 [ 3 2 1]} & 
{\small 35} & {\small -10.388 } & {\small -1/2 } & {\small 1/2 [ 5 4 1]} & 
{\small 57} & {\small -2.261 } & {\small 11/2 } & {\small 11/2 [ 6 1 5]} \\ 
\hline
{\small 14} & {\small -21.292 } & {\small -5/2 } & {\small 5/2 [ 3 1 2]} & 
{\small 36} & {\small -10.280 } & {\small -7/2 } & {\small 7/2 [ 5 2 3]} & 
{\small 58} & {\small -2.207 } & {\small -1/2 } & {\small 1/2 [ 7 5 0]} \\ 
\hline
{\small 15} & {\small -20.333 } & {\small -7/2 } & {\small 7/2 [ 3 0 3]} & 
{\small 37} & {\small -9.794 } & {\small \ \ 1/2 } & {\small 1/2 [ 4 1 1]} & 
{\small 59} & {\small -2.182 } & {\small \ 7/2 } & {\small 7/2 [ 5 0 3]} \\ 
\hline
{\small 16} & {\small -19.621 } & {\small -3/2 } & {\small 3/2 [ 3 1 2]} & 
{\small 38} & {\small -9.301 } & {\small \ \ 5/2 } & {\small 5/2 [ 4 0 2]} & 
{\small 60} & {\small -1.773 } & {\small \ 3/2 } & {\small 3/2 [ 6 4 2]} \\ 
\hline
{\small 17} & {\small -19.254 } & {\small \ 1/2 } & {\small 1/2 [ 4 2 0]} & 
{\small 39} & {\small -9.054 } & {\small -9/2 } & {\small 9/2 [ 5 1 4]} & 
{\small 61} & {\small -1.669 } & {\small \ 1/2 } & {\small 1/2 [ 6 4 0]} \\ 
\hline
{\small 18} & {\small -18.229} & {\small -5/2 } & {\small 5/2 [ 3 0 3]} & 
{\small 40} & {\small -8.356 } & {\small -3/2 } & {\small 3/2 [ 5 3 2]} & 
{\small 62} & {\small -1.553 } & {\small -3/2 } & {\small 3/2 [ 7 4 1]} \\ 
\hline
{\small 19} & {\small -18.189 } & {\small \ \ 3/2 } & {\small 3/2 [ 4 3 1]}
& {\small 41} & {\small -8.276 } & {\small \ \ 1/2 } & {\small 1/2 [ 6 4 0]}
& {\small 63} & {\small -1.459 } & {\small 13/2 } & {\small 13/2 [ 6 0 6]}
\\ \hline
{\small 20} & {\small -18.130 } & {\small -1/2 } & {\small 1/2 [ 3 1 0]} & 
{\small 42} & {\small -8.217 } & {\small -11/2 } & {\small 11/2 [ 5 0 5]} & 
{\small 64} & {\small -1.118 } & {\small -3/2 } & {\small 3/2 [ 5 1 2]} \\ 
\hline
{\small 21} & {\small -16.730 } & {\small \ \ 5/2 } & {\small 5/2 [ 4 2 2]}
& {\small 43} & {\small -7.624 } & {\small -1/2 } & {\small 1/2 [ 5 3 0]} & 
{\small 65} & {\small -1.065 } & {\small -1/2 } & {\small 1/2 [ 5 1 0]} \\ 
\hline
{\small 22} & {\small -16.416 } & {\small \ \ 1/2 } & {\small 1/2 [ 4 3 1]}
& {\small 44} & {\small -7.597} & {\small \ 3/2 } & {\small 3/2 [ 6 5 1]} & 
{\small 66} & {\small -0.393 } & {\small -5/2 } & {\small 5/2 [ 7 5 2]} \\ 
\hline
\end{tabular}
} 
\]

\newpage

\section{Table 2}

\[
\text{%
\begin{tabular}{|l|l|l|l|l|l|l|l|l|l|l|l|}
\hline
{\small \ n} & {\small \ \ [MeV]} & {\small \ \ }$\pi ${\small J} & $\Omega $%
{\small [N n}$_{z}${\small \ }$\Lambda ${\small ]} & n & {\small \ [MeV]} & 
{\small \ \ }$\pi ${\small J} & $\Omega ${\small [N n}$_{z}${\small \ }$%
\Lambda ${\small ]} & n & {\small [MeV]} & {\small \ \ }$\pi ${\small J} & $%
\Omega ${\small [N n}$_{z}${\small \ }$\Lambda ${\small ]} \\ \hline
{\small 1} & {\small -42.0542} & {\small 1/2} & {\small 1/2 [ 0 0 0]} & 
{\small 31} & {\small -19.7288} & {\small -3/2} & {\small 3/2 [ 5 4 1]} & 
{\small 61} & {\small -9.3350} & {\small 1/2} & {\small 1/2 [ 6 4 0]} \\ 
\hline
{\small 2} & {\small -39.0049} & {\small -1/2} & {\small 1/2 [ 1 1 0]} & 
{\small 32} & {\small -18.9901} & {\small 7/2} & {\small 7/2 [ 4 0 4]} & 
{\small 62} & {\small -9.20000} & {\small 11/2 } & {\small 11/2 [ 6 1 5]} \\ 
\hline
{\small 3} & {\small -37.9568} & {\small 3/2} & {\small 3/2 [ 1 0 1]} & 
{\small 33} & {\small -18.9635} & {\small 3/2} & {\small 3/2 [ 4 1 1]} & 
{\small 63} & {\small -8.6428} & {\small -1/2} & {\small 1/2 [ 7 5 0]} \\ 
\hline
{\small 4} & {\small -37.6894} & {\small -1/2} & {\small 1/2 [ 1 0 1]} & 
{\small 34} & {\small -18.4784} & {\small -5/2} & {\small 5/2 [ 5 3 2]} & 
{\small 64} & {\small 8.4314} & {\small 13/2} & {\small 13/2 [ 6 0 6]} \\ 
\hline
{\small 5} & {\small -35.3282} & {\small 1/2} & {\small 1/2 [ 2 2 0]} & 
{\small 35} & {\small -18.1710} & {\small 1/2} & {\small 1/2 [ 4 1 1]} & 
{\small 65} & {\small -8.3129} & {\small -3/2} & {\small 3/2 [ 5 0 1]} \\ 
\hline
{\small 6} & {\small -34.0676} & {\small 3/2} & {\small 3/2 [ 2 1 1]} & 
{\small 36} & {\small -17.7626} & {\small -1/2} & {\small 1/2 [ 5 4 1]} & 
{\small 66} & {\small -8.2710} & {\small -5/2 } & {\small 5/2 [ 5 0 3]} \\ 
\hline
{\small 7} & {\small -33.5860} & {\small 1/2} & {\small 1/2 [ 2 1 1]} & 
{\small 37} & {\small -17.6098} & {\small 5/2 } & {\small 5/2 [ 4 0 2]} & 
{\small 67} & {\small -7.9719} & {\small -3/2 } & {\small 3/2 [ 7 6 1]} \\ 
\hline
{\small 8} & {\small -33.1312} & {\small 5/2} & {\small 5/2 [ 2 0 2]} & 
{\small 38} & {\small -17.1797} & {\small -7/2} & {\small 7/2 [ 5 2 3]} & 
{\small 68} & {\small \ -7.6257} & {\small -1/2 } & {\small 1/2 [ 5 0 1]} \\ 
\hline
{\small 9} & -{\small 32.3233} & {\small 3/2} & {\small 3/2 [ 3 2 1]} & 
{\small 39} & {\small -16.2814} & {\small 3/2} & {\small 3/2 [ 4 0 2]} & 
{\small 69} & {\small -7.4628} & {\small 5/2} & {\small 5/2 [ 6 3 3]} \\ 
\hline
{\small 10} & {\small -31.2577} & {\small 1/2} & {\small 1/2 [ 2 0 0]} & 
{\small 40} & {\small -16.0693} & {\small -9/2} & {\small 9/2 [ 5 1 4]} & 
{\small 70} & {\small -7.4347} & {\small 3/2 } & {\small 3/2 [ 6 3 1]} \\ 
\hline
{\small 11} & {\small -30.9731} & {\small -1/2} & {\small 1/2 [ 3 3 0]} & 
{\small 41} & {\small -15.9987} & {\small 1/2} & {\small 1/2 [ 4 0 0]} & 
{\small 71} & {\small -6.8936} & {\small -5/2} & {\small 5/2 [ 7 5 2]} \\ 
\hline
{\small 12} & {\small -29.7758} & {\small -3/2} & {\small 3/2 [ 3 2 1]} & 
{\small 42} & {\small -15.9048} & {\small -3/2} & {\small 3/2 [ 5 3 2]} & 
{\small 72} & {\small -6.1818} & {\small 1/2} & {\small 1/2 [ 6 3 1]} \\ 
\hline
{\small 13} & {\small -28.9791} & {\small -1/2} & {\small 1/2 [ 3 2 1]} & 
{\small 43} & {\small -15.3084} & {\small -1/2} & {\small 1/2 [ 5 3 0]} & 
{\small 73} & {\small -5.6980} & {\small -5/2} & {\small 5/2 [ 6 2 2]} \\ 
\hline
{\small 14} & {\small -28.5603} & {\small -5/2} & {\small 5/2 [ 3 1 2]} & 
{\small 44} & {\small -15.2725} & {\small -11/2} & {\small 11/2 [ 5 0 5]} & 
{\small 74} & {\small -5.5954} & {\small -7/2 } & {\small 7/2 [ 7 4 3]} \\ 
\hline
{\small 15} & {\small -27.6884} & {\small -7/2 } & {\small 7/2 [ 3 0 3]} & 
{\small 45} & {\small -14.7923} & {\small 1/2} & {\small 1/2 [ 6 4 0]} & 
{\small 75} & {\small -5.5815} & {\small 7/2 } & {\small 7/2 [ 6 2 4]} \\ 
\hline
{\small 16} & {\small -27.2878} & {\small -3/2 } & {\small 3/2 [ 3 1 2]} & 
{\small 46} & {\small -14.0318} & {\small 3/2 } & {\small 3/2 [ 6 5 1]} & 
{\small 76} & {\small -4.4142} & {\small \ -1/2 } & {\small 1/2 [ 7 7 0]} \\ 
\hline
{\small 17} & {\small -26.2258} & {\small -1/2} & {\small 1/2 [ 3 1 0]} & 
{\small 47} & {\small -14.0081} & {\small -5/2 } & {\small 5/2 [ 5 2 3]} & 
{\small 77} & {\small -4.3051} & {\small 7/2} & {\small 7/2 [ 6 1 3]} \\ 
\hline
{\small 18} & {\small -26.0583} & {\small -5/2} & {\small 5/2 [ 3 0 3]} & 
{\small 48} & {\small -13.2796} & {\small -3/2 } & {\small 3/2 [ 5 2 1]} & 
{\small 78} & {\small -4.2459} & {\small -9/2} & {\small 9/2 [ 7 3 4]} \\ 
\hline
{\small 19} & {\small -26.0283} & {\small 1/2} & {\small 1/2 [ 4 4 0]} & 
{\small 49} & {\small -12.8670} & {\small 5/2} & {\small 5/2 [ 6 4 2]} & 
{\small 79} & {\small -4.0132} & {\small 9/2} & {\small 9/2 [ 6 1 5]} \\ 
\hline
{\small 20} & {\small -24.9930} & {\small 3/2 } & {\small 3/2 [ 4 3 1]} & 
{\small 50} & {\small -12.3859} & {\small -7/2} & {\small 7/2 [ 5 1 4]} & 
{\small 80} & {\small -3.9400} & {\small 1/2 } & {\small 1/2 [ 6 2 0]} \\ 
\hline
{\small 21} & {\small -24.5673 } & {\small -3/2} & {\small 3/2 [ 3 0 1]} & 
{\small 51} & {\small -12.2367} & {\small -1/2} & {\small 1/2 [ 5 2 1]} & 
{\small 81} & {\small -3.7016} & {\small 3/2} & {\small 3/2 [ 6 2 2]} \\ 
\hline
{\small 22} & {\small -24.0164} & {\small -1/2} & {\small 1/2 [ 3 0 1]} & 
{\small 52} & {\small -11.6318} & {\small -5/2 } & {\small 5/2 [ 5 1 2]} & 
{\small 82} & {\small -3.3495} & {\small 9/2} & {\small 9/2 [ 6 0 4]} \\ 
\hline
\end{tabular}
} 
\]

\bigskip 
\[
\begin{tabular}{|l|l|l|l|l|l|l|l|l|l|l|l|}
\hline
{\small 23} & {\small -23.7133} & {\small 1/2} & {\small 1/2 [ 4 3 1]} & 
{\small 53} & {\small -11.5394} & {\small 7/2} & {\small 7/2 [ 6 3 3]} & 
{\small 83} & {\small -3.0797} & {\small -1/2} & {\small 1/2 [ 7 6 0]} \\ 
\hline
{\small 24} & {\small -23.7025} & {\small 5/2} & {\small 5/2 [ 4 2 2]} & 
{\small 54} & {\small -11.2224} & {\small 1/2} & {\small 1/2 [ 6 5 1]} & 
{\small 84} & {\small -3.0058} & {\small -11/2} & {\small 11/2 [ 7 2 5]} \\ 
\hline
{\small 25} & {\small -22.5412} & {\small 7/2} & {\small 7/2 [ 4 0 4]} & 
{\small 55} & {\small -11.2192} & {\small -9/2 } & {\small 9/2 [ 5 0 5]} & 
{\small 85} & {\small -2.8563} & {\small 11/2} & {\small 11/2 [ 6 0 6]} \\ 
\hline
{\small 26} & {\small -21.8511} & {\small 3/2} & {\small 3/2 [ 4 2 2]} & 
{\small 56} & {\small 10.4881} & {\small -7/2} & {\small 7/2 [ 5 0 3]} & 
{\small 86} & {\small -2.6835} & {\small -3/2} & {\small 3/2 [ 7 6 2]} \\ 
\hline
{\small 27} & {\small -21.7120} & {\small 9/2} & {\small 9/2 [ 4 0 4]} & 
{\small 57} & {\small -10.2634} & {\small 9/2 } & {\small 9/2 [ 6 2 4]} & 
{\small 87} & {\small -2.3053} & {\small 3/2} & {\small 3/2 [ 6 1 1]} \\ 
\hline
{\small 28} & {\small -20.9447} & {\small 1/2} & {\small 1/2 [ 4 2 0]} & 
{\small 58} & {\small -9.8951} & {\small -3/2 } & {\small 3/2 [ 5 1 2]} & 
{\small 88} & {\small -2.2208} & {\small 1/2} & {\small 1/2 [ 8 6 0]} \\ 
\hline
{\small 29} & {\small -20.6103} & {\small -1/2} & {\small 1/2 [ 5 5 0]} & 
{\small 59} & {\small -9.8517} & {\small -1/2 } & {\small 1/2 [ 5 1 0]} & 
{\small 89} & {\small -1.9882} & {\small -13/2} & {\small 13/2 [ 7 1 6]} \\ 
\hline
{\small 30} & {\small -20.1878} & {\small 5/2} & {\small 5/2 [ 4 1 3]} & 
{\small 60} & {\small -9.4457} & {\small 3/2} & {\small 3/2 [ 6 4 2]} & 
{\small 90} & {\small -1.8461} & {\small 5/2} & {\small 5/2 [ 6 1 3]} \\ 
\hline
\end{tabular}
\]

\newpage

\section{Figure Captions}

\bigskip

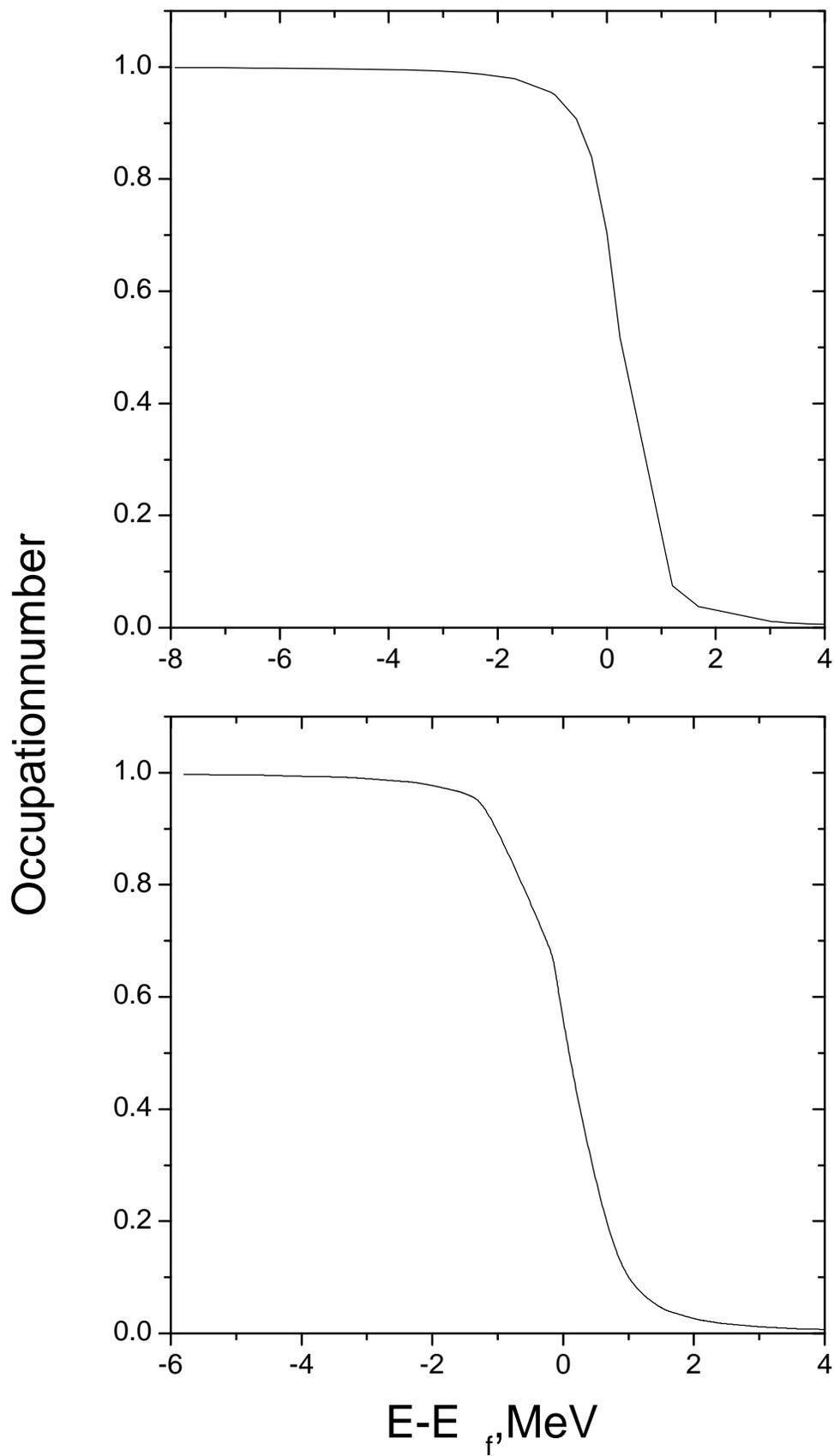
{\bf Fig.1}. Occupation probabilities for the single particle bound states
of $\ ^{238}$U. Upper part corresponds to protons, lower to neutrons.

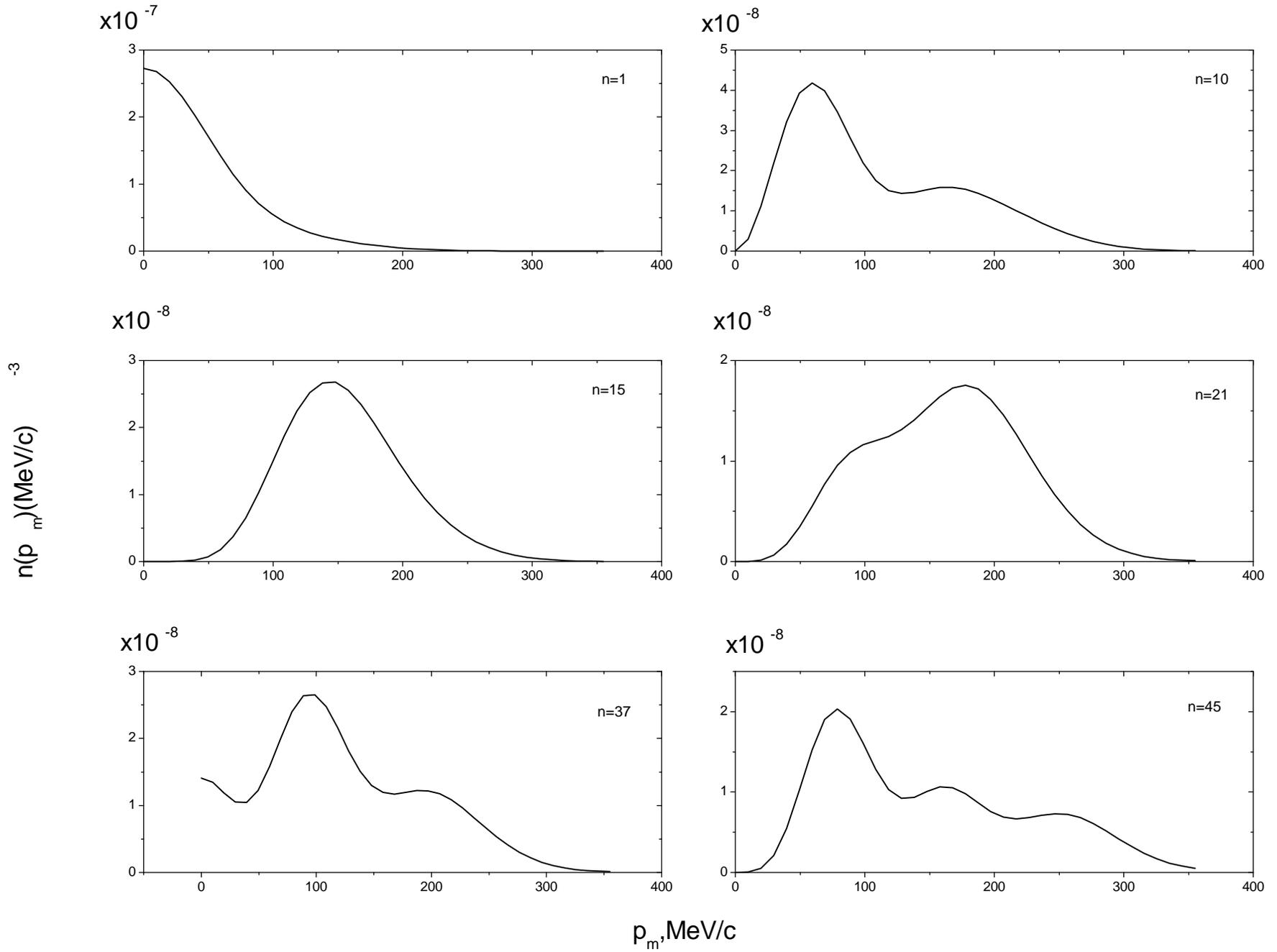
{\bf Fig.2}. Typical momentum distribution for proton bound states.

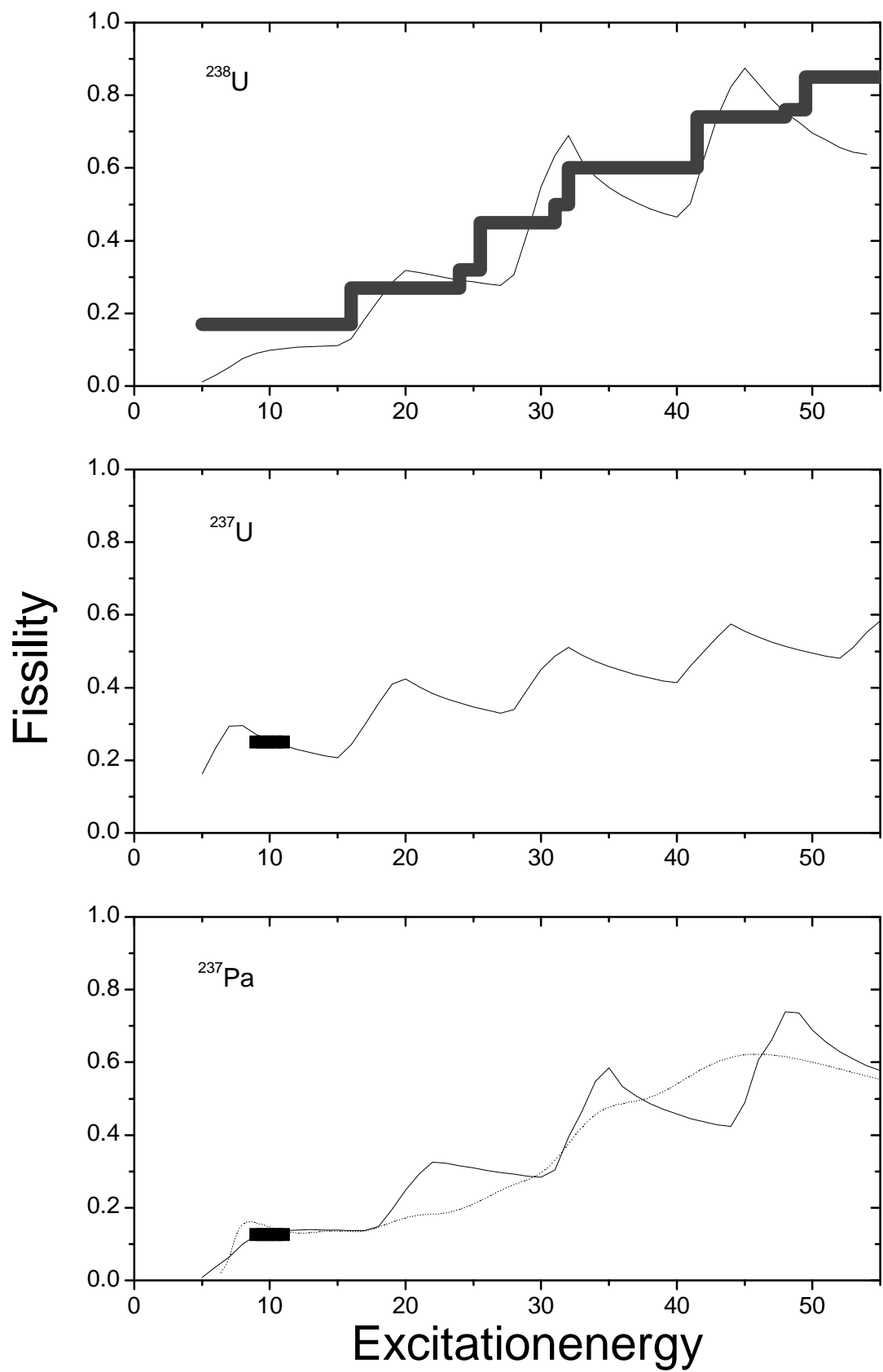
{\bf Fig.3}. Fissility for $^{237}Pa$ and$^{\text{ \ \ \ }237,238}$U. The
solid (dotted) curve shows the results of calculations on the statistical
approach without (with) account of preequilibrium particle emission. The
dark strips show the experimental data \ (see text for details).

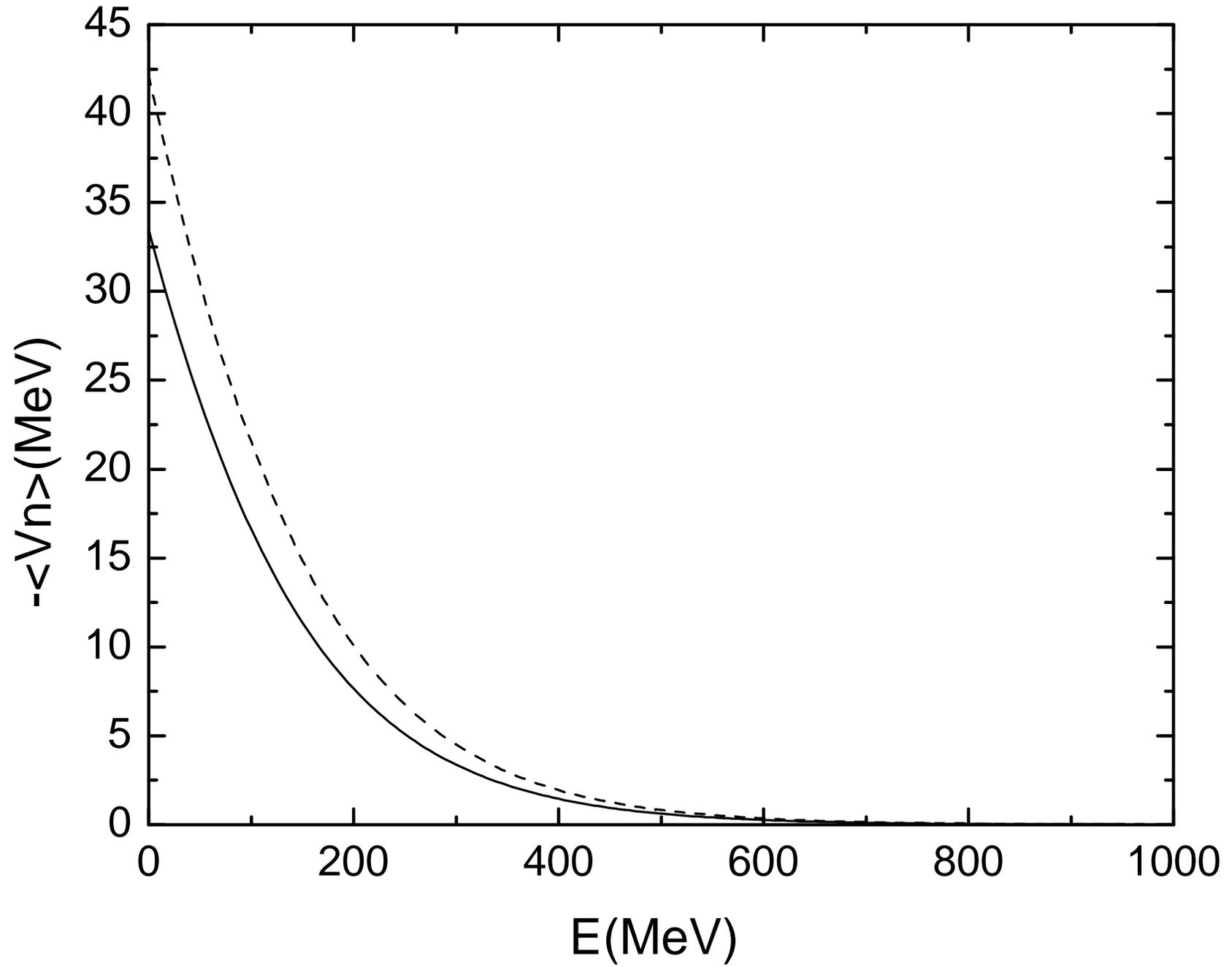
{\bf Fig.4}. Average real potential depth as a function of the \ nucleon
energy. \ The dashed curve corresponds to protons (residual nucleus $%
^{237}Pa);$ solid curve to neutrons ($^{237}U).$

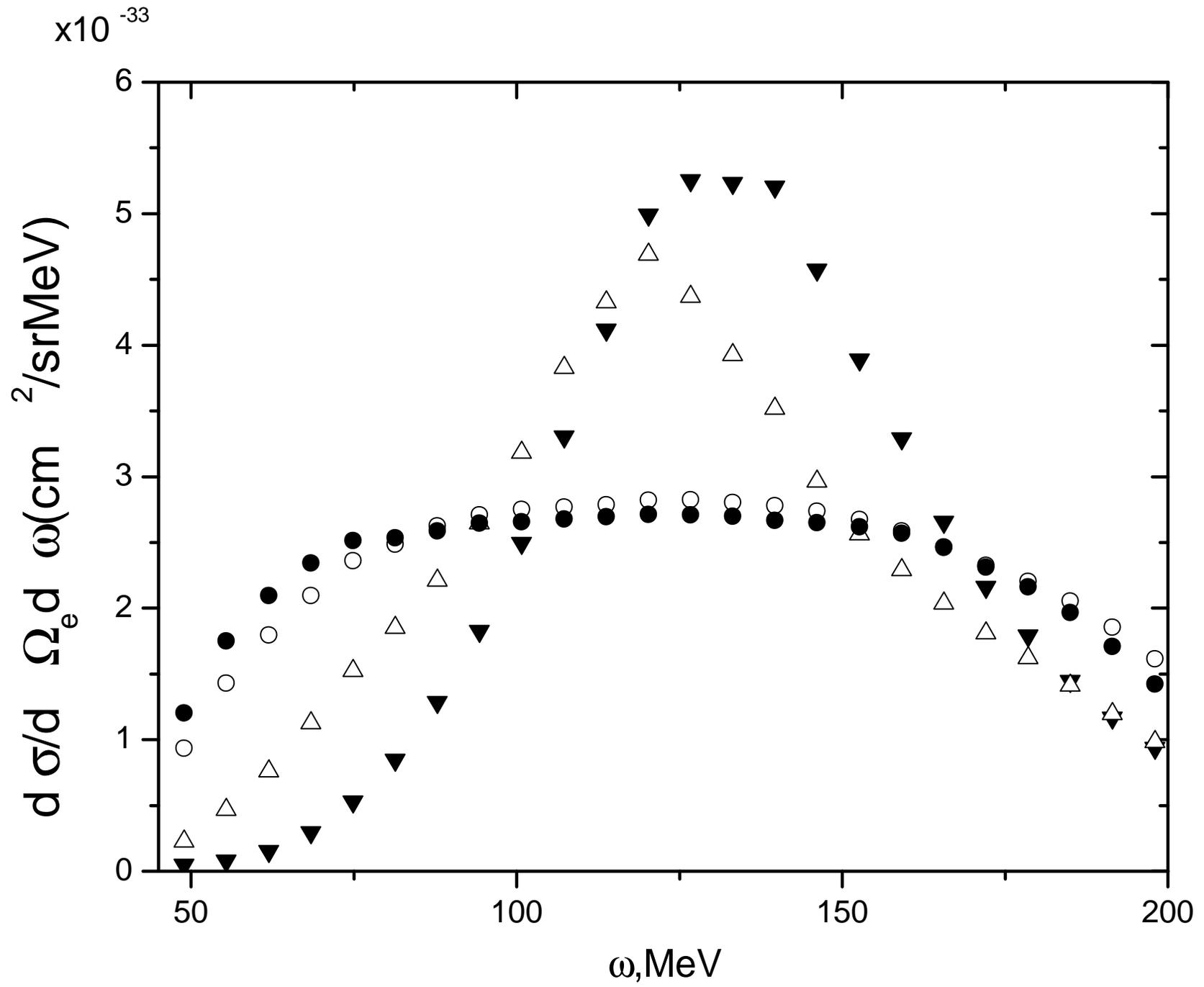
{\bf Fig.5}. PWIA quasi free $(e,e^{\prime })-$\ cross section for four
proton orbitals of $^{238}$U and $\varepsilon _{1}=750$ MeV, $\theta
e=37.5^{0}$. The dark triangles correspond to n=1 (see table 1), light
triangles to n=11, light circles to n=21, and dark circles to n=31.

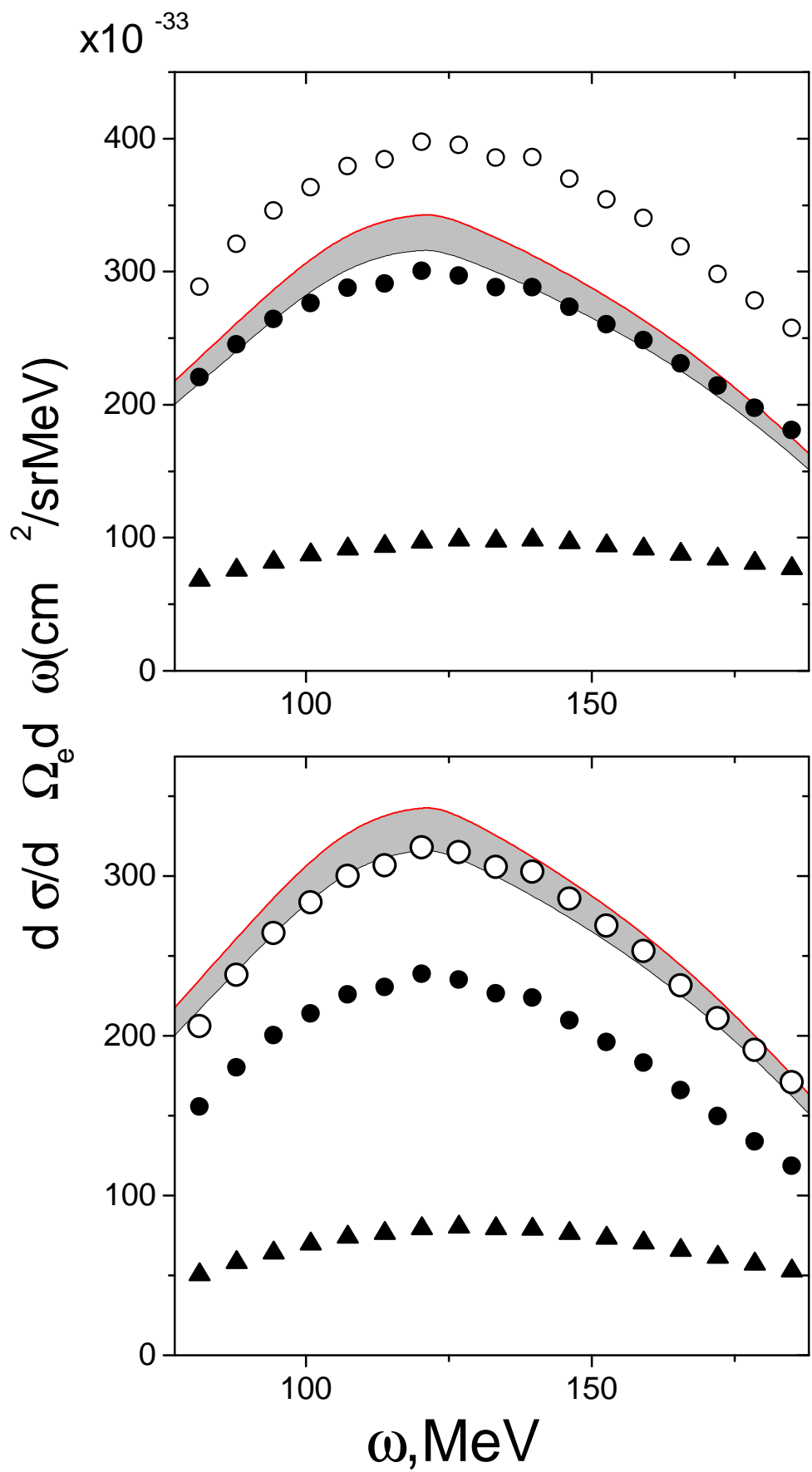
{\bf Fig.6}. Quasi free $(e,e^{\prime })-$cross section for $\varepsilon
_{1}=750$ MeV and $\theta e=37.5^{0}$. The upper part shows the results of
calculations in \ PWIA, lower -with account of distortion corrections (see
text for details) . The dark circles show the proton contribution, dark
triangles - neutron, and light circles - total cross section. The hatched
area represents the experimental data \cite{Hansen}

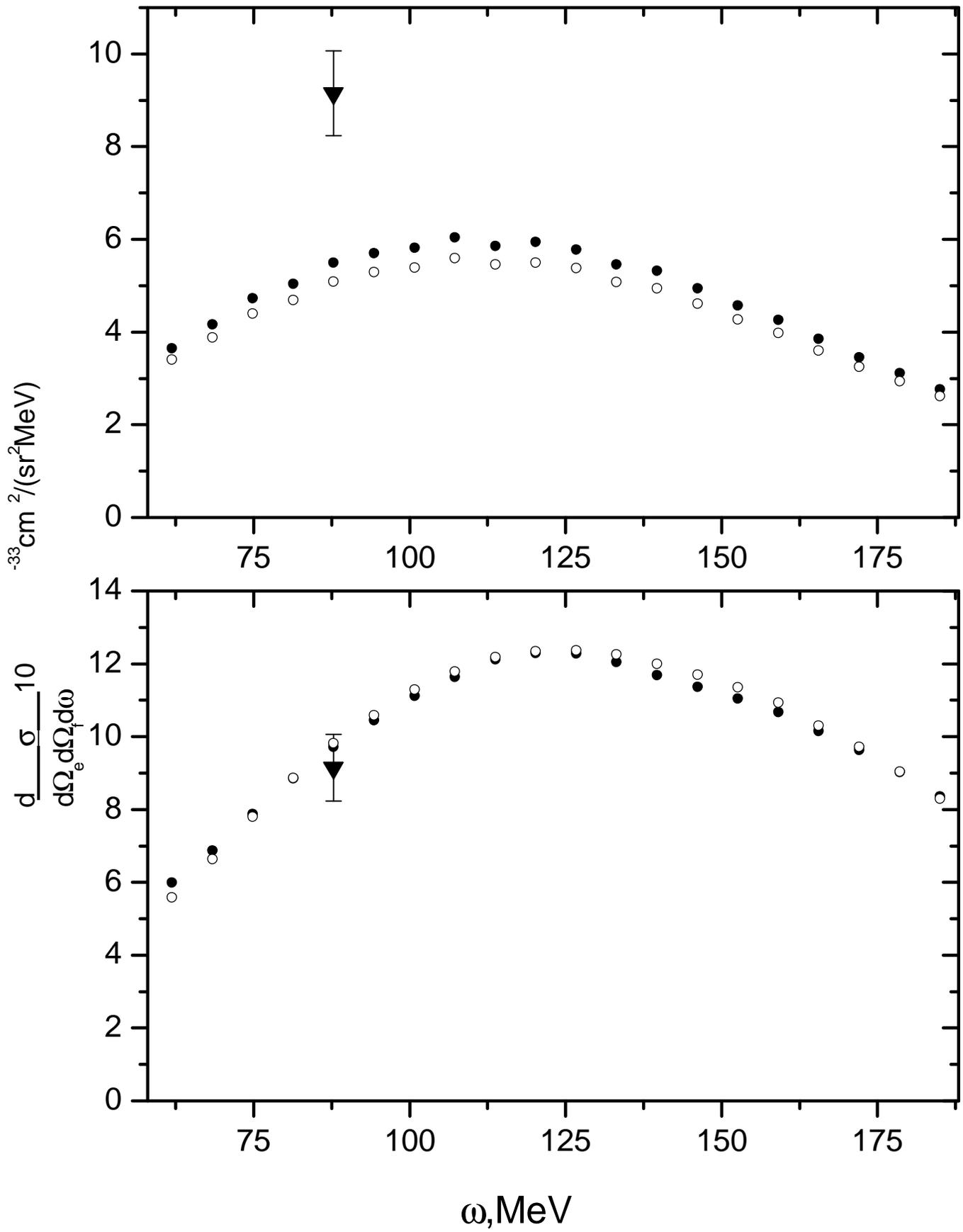
{\bf Fig. 7}. Total quasi free $(e,e^{\prime }f)-$cross section for $%
\varepsilon _{1}=750$ MeV and $\ \theta e=37.5^{0}$ . The upper part shows \
the calculation in PWIA with account of distortion corrections to the cross
section, but without account of FSI corrections \ to the fissility (see text
for details). The lower part shows results similar to the upper part, but
with account of FSI corrections \ to the fissility. The light circles show
the results of calculations which take into account the preequilibrium
emission, and dark circles - without accounting of\ the preequilibrium
emission. The dark triangle with error bars represents the experimental data 
\cite{Hansen}

\newpage

\section{Table Caption}

\bigskip

\ \ \ {\bf Table 1}.\ Proton single-particle levels of $\ ^{238}$U. The
Fermi level is the level 46.

\bigskip

\ \ \ {\bf Table 2}.\ Neutron single-particle levels of $\ ^{238}$U. The
Fermi level is the level 73.

\end{document}